\documentclass[showpacs,aps,graphicx,twocolumn]{revtex4}

\usepackage{graphicx}

\begin{document}

\title{Selective-Resonance-Based  Quantum  Entangling  Operation  on  Qubits in  Circuit QED}

\author{Ming Hua and Fu-Guo Deng\footnote{Corresponding author: fgdeng@bnu.edu.cn}}

\address{Department of  Physics, Applied Optics Beijing Area Major Laboratory, Beijing
Normal University, Beijing 100875, China}

\date{\today }

\begin{abstract}
We present a fast quantum entangling operation on superconducting
qubits assisted by a resonator in the quasi-dispersive regime with a
new effect --- the selective resonance coming from the amplified
qubit-state-dependent resonator transition frequency and the tunable
period relation between a wanted quantum Rabi oscillation and an
unwanted one. This operation does not require any kind of drive
fields and the interaction between qubits.  More interestingly, the
non-computational third excitation states of the charge qubits can
play an important role in shortening  largely the operation time of
the entangling gates. All those features provide an efficient way to
realize much faster quantum entangling gates on superconducting
qubits than previous proposals.
\end{abstract}

 \maketitle

Quantum information and quantum computation \cite{Nielsen}  has
attracted much attention. Constructing universal quantum gates and
generating multipartite entanglement are two key tasks in this
topic, especially those based on superconducting qubits
\cite{squbit1,squbit2,squbit3,squbit4}. Circuit quantum
electrodynamics (QED), combining the method of cavity QED and
superconducting circuits, has been widely studied for quantum
information  processing
\cite{Wallraff,JQYou,DiCarlo,Lucero,reed,liu1,Rigetti,Jerry2011,Jerry,natureletter},
because of both its long-coherence time and its good scalability.
Since the first physical mode was proposed by Yale group
\cite{Wallraff}, circuit QED has been used for resolving photon
number states in a superconducting circuit \cite{DISchuster},
constructing the quantum non-demolition detector for measuring the
number of photons inside a high-quality-factor microwave cavity on a
chip \cite{Johnson}, simulating the basic interaction between an
atom and a cavity even in the ultrastrong coupling regime
\cite{Ballester}, and realizing the quantum information processing
between superconducting qubits or between microwave photons
\cite{DiCarlo,reed,Lucero,JQYou,FrederickWStrauch,frederick,han}.

Circuit QED can also provide  an effective control on
superconducting qubits in some important regimes  by choosing  the
coupling strength between a qubit and a cavity \cite{DISchuster},
such as the dispersive regime ($\vert
\frac{g_{i}^{j}}{\Delta_{i}^{j} }\vert <<1$) \cite{Alexandre,Hut},
the quasi-dispersive regime ($0.1<\vert
\frac{g_{i}^{j}}{\Delta_{i}^{j} }\vert <1$)
 \cite{DISchuster,frederick}, the resonant regime ($\vert
\frac{g_{i}^{j}}{\Delta_{i}^{j} }\vert >1$)
 \cite{Sillanpaa,FrederickWStrauch}, and even the ultrastrong
coupling regime ($0.1<\vert \frac{g_{i}^{j}}{\omega _{r_{j}} }\vert
<1$)  \cite{Ballester,Romero}. Here $g_{i}^{j}$ is the coupling
strength between the qubit $q_{i}$ and the resonator $R_{j}$.
$\omega _{r_{j}}$ and $\omega _{i}$ are the frequencies of $ R_{j}$
and $q_{i}$, respectively. $\Delta_{i}^{j} =\omega _{r_{j}}-\omega
_{i} $. Aiming to find an effective high-fidelity quantum entangling
operation on superconducting qubits in circuit QED with shorter
operation time, we pay close attention to two of the critical
characters of the dispersive regime in cavity QED
\cite{FrederickWStrauch}: the qubit-state-dependent (QSD) transition
  on a cavity
 \cite{Filipp,chow} and the number-state-dependent (NSD) transition  on a
qubit  \cite{frederick,Gambetta}. By increasing the coupling
strength between a qubit and a cavity, the QSD resonator transition
frequency and the NSD qubit transition frequency can be amplified
effectively. The amplified NSD qubit transition frequency in circuit
QED has been studied by Strauch, Jacobs, and Simmonds in 2010, and
they gave an interesting effect --- a selective rotation
\cite{frederick} with a drive field, which can be used to generate
the entanglement between two microwave photons effectively.

In this paper, we investigate the effect from the amplified QSD
resonator transition frequency in the quasi-dispersive regime in
circuit QED and its application in quantum entangling operation and
entanglement generation for superconducting qubits. We use charge
superconducting qubits to describe our results  by taking the
influence from the non-computational third excitation states of the
qubits into account.

\begin{figure}[tpb]
\begin{center}
\includegraphics[width=7.8 cm,angle=0]{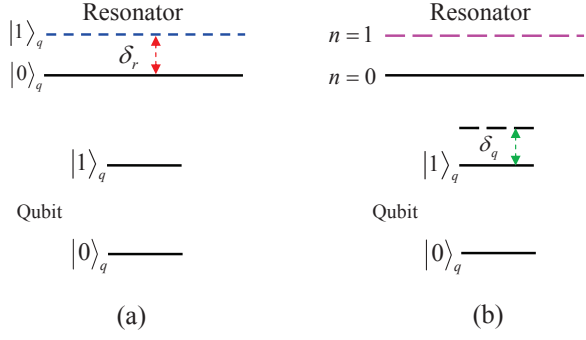} 
\caption{(color online) (a) The  qubit-state-dependent  resonator
transition, which means the frequency shift of the resonator
transition $\delta_{r} $ arises from the state ($\vert 0\rangle_{q}$
or $\vert 1\rangle_{q}$) of the qubit. (b) The
number-state-dependent qubit transition, which means the frequency
shift $\delta_{q} $ takes place on the qubit due to the photon
number $n=1$ or $0$ in the resonator in the dispersive regime.}
\label{fig1}
\end{center}
\end{figure}

The Hamiltonian of  a  superconducting  qubit  coupled to a
resonator can be described as ($\hbar =1$),
\begin{equation}    
H=\omega _{r}a^{+}a+\omega \sigma ^{+}\sigma ^{-}+g\left( a+a^{+}\right)
(\sigma ^{-}+\sigma ^{+}),  \label{jaynescummings}
\end{equation}
where $\sigma ^{+}=\vert 1\rangle \langle 0\vert $ and $a^{+}$ are
the creation operators of a qubit $q$ and a resonator $R$,
respectively. When we consider the dispersive regime, after making rotating-wave approximation and a
unitary transformation $U=\exp \left[ \frac{g}{\Delta }\left(a\sigma
^{+}-a^{+}\sigma ^{-}\right)\right]$ on the Hamiltonian $ H $,
we can get
\begin{equation} 
UHU^{+}\approx \omega _{r}a^{+}a+\frac{1}{2}\left[\omega
_{q}+\frac{g^{2}}{ \Delta }(2a^{+}a+1)\right]\sigma _{z}
\label{stark}
\end{equation}
or
\begin{equation}  
UHU^{+}\approx \left(\omega _{r}+\frac{g^{2}}{\Delta }\sigma
_{z}\right)a^{+}a+ \frac{1}{2}\left(\omega _{q}+\frac{g^{2}}{\Delta
}\right)\sigma _{z}.  \label{kerr}
\end{equation}
Eq.(\ref{stark}) means  the NSD qubit transition  and Eq.(\ref
{kerr}) means  the QSD resonator transition, shown in
Fig.\ref{fig1}(a) and (b), respectively.

The NSD qubit transition comes from the effect that the transition
frequency of the qubit depends on the photon number in the resonator
with $ \omega _{q}^{\prime }=\omega _{q}+\frac{g^{2}}{\Delta
}(2a^{+}a+1)$, where $ \omega _{q}^{\prime }$ is the frequency of
the qubit after the NSD effect. The QSD resonator transition is the
effect that the transition frequency of the resonator depends on the
state of the qubit with $\omega _{r}^{\prime }=\omega _{r}+\sigma
_{z}\frac{g^{2}}{\Delta }$. Here $\omega _{r}^{\prime }$ is the
frequency of the resonator after the QSD effect. From
Eq.(\ref{stark}) and Eq.(\ref{kerr}), one can see that the NSD qubit
transition frequency and the QSD resonator transition frequency can
be amplified effectively when $\frac{g^{2}}{\Delta }$ is large
enough, which means the number of photons in the cavity can make a
large shift on the transition frequency of the qubit and the qubit
in different states will also make a large shift on the transition
frequency of the resonator. The NSD qubit transition in the
quasi-dispersive regime was used as an effective method to realize
entanglement and quantum gates between microwave photons
\cite{frederick,Johnson,DISchuster}. In fact, the QSD resonator
transition frequency can also be amplified in the quasi-dispersive
regime effectively.

\begin{figure}[tpb]
\begin{center}
\includegraphics[width=8 cm,angle=0]{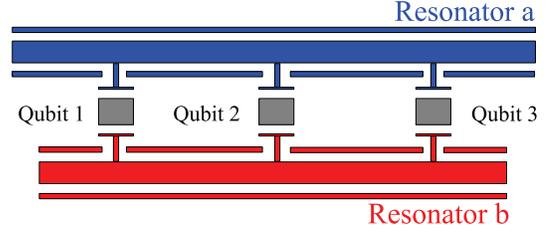} 
\caption{(color online) Sketch of a coplanar geometry for the
circuit QED with three superconducting qubits.  Qubits are placed
around the maxima of the electrical field amplitude of $R_{a}$ and
$R_{b}$ (not drawn in this figure), and the distance between them is
large enough so that there is no direct interaction between them.
The fundamental frequencies of resonators are $\omega _{r_{j}}/(2\pi
)$ ($j=a,b$), the frequencies of the qubits are $\omega
_{q_{i}}/(2\pi )$ ($i=1,2,3$), and they are capacitively coupled to
the resonators. The coupling strengths between them are
$g_{i}^{j}/(2\pi )$. We can use the control line (not drawn here) to
afford the flux to tune the transition frequencies of the qubits.}
\label{fig2}
\end{center}
\end{figure}

\begin{figure}[tpb]
\begin{center}
\includegraphics[width=8.0 cm,angle=0]{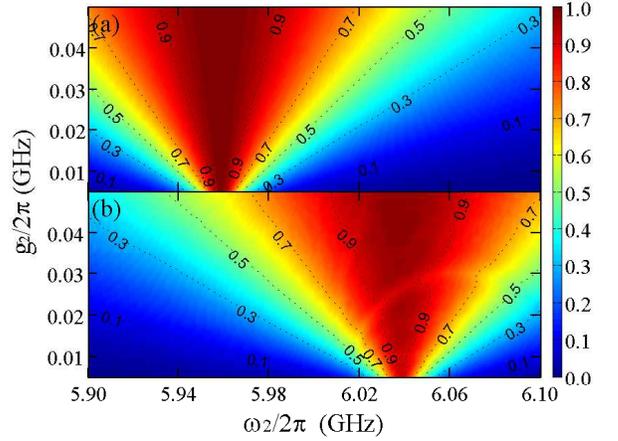} 
\caption{(color online) Simulated outcomes for the maximum amplitude
value of the expectation about the quantum Rabi oscillation varying
with the coupling strength $g_{2}$ and the frequency of the second
qubit $\omega _2$. (a) The outcomes for ROT$_0$: $\vert 0\rangle
_{1}\vert 1\rangle _{2}\vert 0\rangle _{a} \leftrightarrow \vert
0\rangle _{1}\vert 0\rangle _{2}\vert 1\rangle _{a}$. (b) The
outcomes for ROT$_1$: $\vert 1\rangle _{1}\vert 1\rangle _{2}\vert
0\rangle _{a}\leftrightarrow \vert 1\rangle _{1}\vert 0\rangle
_{2}\vert 1\rangle _{a}$. Here the parameters of the resonator and
the first qubit $q_{1}$ are taken as $\omega _{a}/(2\pi )=6.0$ GHz,
$\omega _{q_{1}}/(2\pi )=7.0$ GHz, and $g_{1}/(2\pi )=0.2$ GHz.}
\label{fig3}
\end{center}
\end{figure}

In order to show the  influence  on the resonance  between a qubit
and a resonator by the  amplified  QSD  resonator transition
frequency, let us consider the case that two perfect  qubits  couple
to a resonator (i.e., $R_{a}$) with the Hamiltonian
\begin{eqnarray}    
H_{2q} &=&\omega _{r_{a}}a^{+}a+\omega _{1}\sigma _{1}^{+}\sigma
_{1}^{-}+\omega _{2}\sigma _{2}^{+}\sigma _{2}^{-}  +g_{1}\left(
a^{+}+a\right)\nonumber \\
&& \otimes (\sigma _{1}^{+}+\sigma _{1}^{-}) +g_{2}\left(
a^{+}+a\right) (\sigma _{2}^{+}+\sigma _{2}^{-}) \label{2q}
\end{eqnarray}
in which we neglect the direct interaction between the two  qubits
(i.e., $q_{1}$ and $q_{2}$), shown in Fig.\ref{fig2}. Here $\sigma
^{+}_i=\vert 1\rangle_i \langle 0\vert $ is the creation operator of
$q_i$ ($i=1,2$). $g_i$ is the coupling strength between $q_i$ and
$R_a$. The parameters are chosen to make $q_{1}$ interact with $R_a$
in the quasi-dispersive regime. That is, the transition frequency of
$R_a$ is determined by the state of $q_{1}$. By taking a proper
transition frequency of $q_{2}$ (which equals to the transition
frequency of $R_a$ when $q_{1}$ is in the state $ \vert 0\rangle
_{1}$), one can realize the quantum Rabi oscillation (ROT) ROT$_0$:
$\vert 0\rangle _{1}\vert 1\rangle _{2}\vert 0\rangle _{a}
\leftrightarrow \vert 0\rangle _{1}\vert 0\rangle _{2}\vert 1\rangle
_{a}$, while ROT$_1$: $\vert 1\rangle _{1}\vert 1\rangle _{2}\vert
0\rangle _{a} \leftrightarrow \vert 1\rangle _{1}\vert 0\rangle
_{2}\vert 1\rangle _{a}$ occurs with a small probability as $q_{2}$
detunes with $R_a$ when $q_{1}$ is in the state $\vert 1\rangle
_{1}$. Here the Fock state $\vert n\rangle _{a}$ represents the
photon number $n$ in $R_{a}$ ($n=0,1$). $ \vert 0\rangle _{i}$ and
$\vert 1\rangle _{i}$ are the ground and the first excited states of
 $q_{i}$, respectively.

We numerically simulate  the maximal expectation values (MAEVs) of
ROT$_0$ and ROT$_1$ based on  the Hamiltonian $H_{2q}$, shown in
Fig. \ref{fig3}(a) and (b), respectively. Here, the expectation
value is defined as $ \vert \langle \psi \vert e^{-iH_{2q}t/\hbar
}\vert \psi _{0}\rangle \vert ^{2}$. $\vert \psi _{0}\rangle$ and
$\vert \psi \rangle$ are the initial and the final states of a
quantum Rabi oscillation, respectively. The  MAEV s vary with the
transition frequency $\omega _2$ and the coupling strength $g_{2}$.
It is obvious that the amplified  QSD  resonator transition can
generate a \emph{selective resonance} (SR) when the coupling
strength $g_{2}$ is small enough.

It is worth noticing that there is a detune  between $q_{2}$ and
$R_a$ in  ROT$_1$, when $q_{2}$ is resonant with $R_a$ for realizing
ROT$_0$ with a large probability.  In principle, the frequency of
the quantum Rabi oscillation between a qubit and a resonator can be
described as  \cite{scully}
\begin{equation}    
\Omega _{n}^{2}=\Delta ^{2}+4g^{2}(n+1),  \label{rabioscillation}
\end{equation}
where $n$ is the number of photons in the resonator. One can see
that the period of  ROT$_0$ is different from that of ROT$_1$ as
$\Delta=0$ for  ROT$_0$ and $\Delta=\delta_r$ for  ROT$_1$. By
taking a proper parameter for $g$, we can tune the different period
relation between these two quantum Rabi oscillations. That is, a
\emph{tunable period relation} between a wanted quantum Rabi
oscillation and an unwanted one can also be obtained.

\begin{figure}[tpb]
\begin{center}
\includegraphics[width=8.0 cm,angle=0]{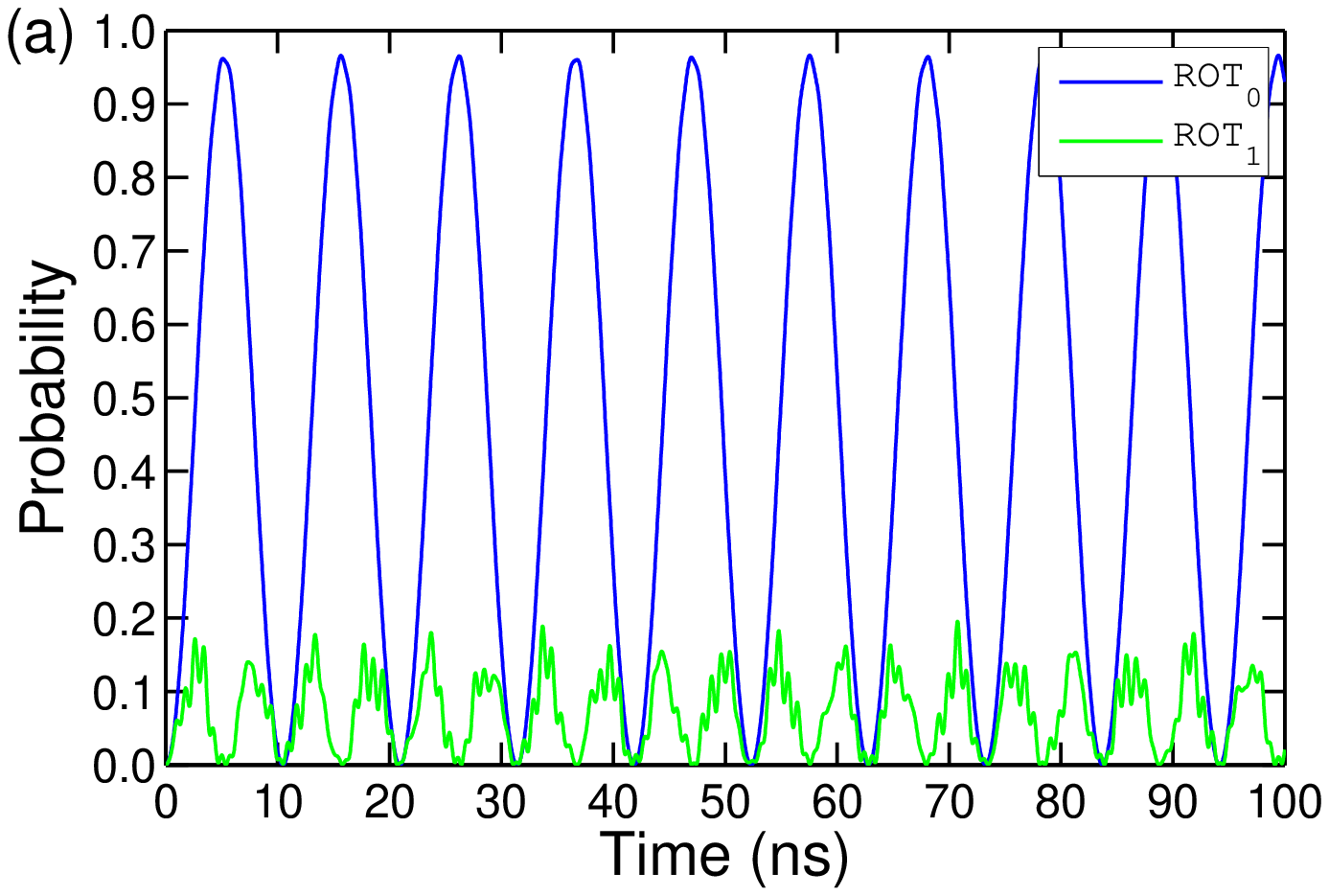} \\ 
\includegraphics[width=8.0 cm,angle=0]{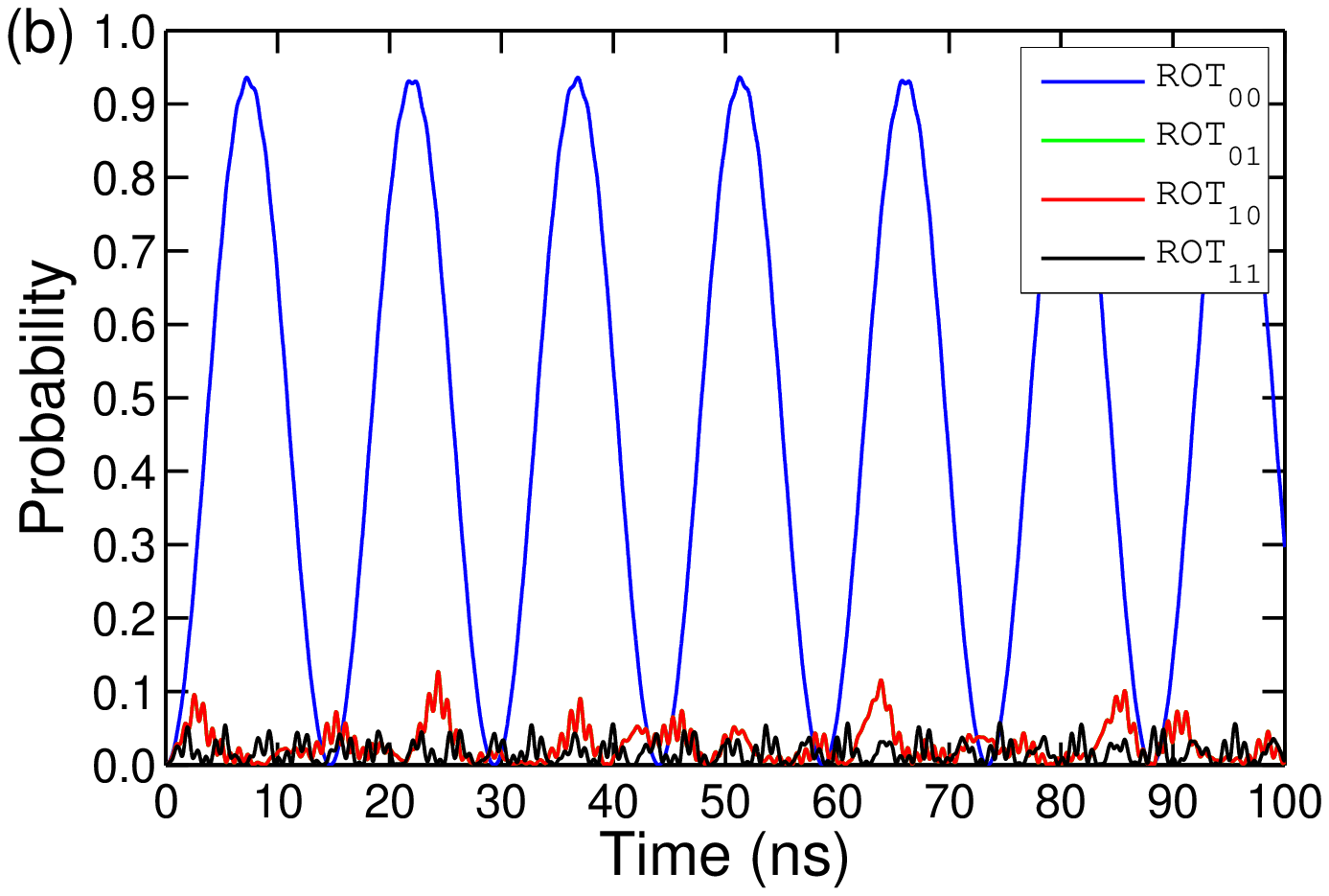} \\ 
\caption{(color online) (a) The probability distribution of the two
quantum Rabi oscillations ROT$_0$ (the blue solid line) and ROT$_1$
(the green solid line) of two charge qubits coupled to a resonator.
(b) The probability distribution of the four quantum Rabi
oscillations in our  cc-phase gate on  a three-charge-qubit system.
Here, the blue-solid, green-solid, red-dashed, and
Cambridge-blue-dot-dashed lines represent the quantum Rabi
oscillations ROT$_{00}$ ($\vert 0\rangle _{1}\vert 0\rangle
_{2}\vert 1\rangle _{3}\vert 0\rangle _{a}\leftrightarrow \vert
0\rangle _{1}\vert 0\rangle _{2}\vert 0\rangle _{3}\vert 1\rangle
_{a}$),  ROT$_{01}$ ($\vert 0\rangle _{1}\vert 1\rangle _{2}\vert
1\rangle _{3}\vert 0\rangle _{a}\leftrightarrow \vert 0\rangle
_{1}\vert 1\rangle _{2}\vert 0\rangle _{3}\vert 1\rangle _{a}$),
ROT$_{10}$ ($\vert 1\rangle _{1}\vert 0\rangle _{2}\vert 1\rangle
_{3}\vert 0\rangle _{a}\leftrightarrow \vert 1\rangle _{1}\vert
0\rangle _{2}\vert 0\rangle _{3}\vert 1\rangle _{a}$), and
ROT$_{11}$ ($\vert 1\rangle _{1}\vert 1\rangle _{2}\vert 1\rangle
_{3} \vert 0\rangle _{a}\leftrightarrow \vert 1\rangle _{1}\vert
1\rangle _{2}\vert 0\rangle _{3}\vert 1\rangle _{a}$),
respectively.} \label{fig4}
\end{center}
\end{figure}

In the discussion below, we consider  practical charge
superconducting qubits in which there are two computational levels
$\vert 0\rangle_q$ and $\vert 1\rangle_q$ for our
selective-resonance-based entangling operation, by taking the
influence from the non-computational third excitation state $\vert
2\rangle_q$ of each charge qubit  into account. When the two charge
qubits $q_1$ and $q_2$ are coupled to the resonator $R_a$, the SR is
simulated with the Hamiltonian
\begin{eqnarray}     
H'_{2q} &\!\!=\!\!& \!\!\!\!\sum\limits_{i=0,1,2\atop q=1,2}\!\!\!\!
E_{i;q}\left\vert i\right\rangle_{q} \! \left\langle
i \right\vert \!+\! \omega _{r_{a}}\! a^{+}\!a \!+\! g_{0\!,1;1}(\!a^{+} \! \!+\! a)(\!\sigma _{0\!,1;1}^{+} \!\!+\! \sigma _{0\!,1;1}^{-})  \nonumber \\
&&+g_{1\!,2;1}(\!a^{+} \!+\! a)(\!\sigma _{1,2;1}^{+} \!+\! \sigma _{1\!,2;1}^{-}\!) \!+\! g_{0\!,1;2}(\!a^{+}\!+\!a)  \nonumber \\
&&\otimes(\!\sigma _{0\!,1;2}^{+} \!+\!\sigma _{0\!,1;2}^{-})\!+\!
g_{1\!,2;2}(\!a^{+}\!+\! a)(\!\sigma _{1\!,2;2}^{+} \!+\! \sigma
_{1\!,2;2}^{-}), \label{threelevel}
\end{eqnarray}
and it is shown in Fig.\ref{fig4} (a). In the simulation of our SR,
we choose  the  reasonable parameters by considering the energy
level structure of a charge qubit, according to Ref. \cite{Koch}.
Here  $\omega _{r_{a}}/(2\pi )=6.0$ GHz. The transition frequency of
two qubits between $\left\vert 0\right\rangle_{q} \leftrightarrow $
$\left\vert 1\right\rangle_{q}$ and $\left\vert 1\right\rangle_{q}
\leftrightarrow $ $\left\vert 2\right\rangle_{q}$ are chosen as
$\omega _{0\!,1;1}/(2\pi )=E_{1;1}-E_{0;1}=5.0$ GHz, $\omega
_{1\!,2;1}/(2\pi )=E_{2;1}-E_{1;1}=6.2$ GHz, $\omega
_{0\!,1;2}/(2\pi )=E_{1;2}-E_{0;2}=6.035$ GHz, and $\omega
_{1\!,2;2}/(2\pi )=E_{2;2}-E_{1;2}=7.335$ GHz.  Here $E_{i;q}$ is
the energy for the level $i$ of the qubit $q$, and $\sigma
_{i,i';q}^{+}\equiv \left\vert i\right\rangle_{q} \left\langle
i'\right\vert$.
 $g_{i,j;q}$ is the coupling strength between the resonator $R_a$ and
the qubit $q$ in the transition between the energy levels $\vert
i\rangle_q$ and $\vert j\rangle_q$ ($i=0,1$, $j=1,2$, and $q=1,2$).
For convenience, we take the coupling strengths as
$g_{0\!,1;1}/(2\pi )=g_{1\!,2;1}/(2\pi )=0.2$ GHz and
$g_{0\!,1;2}/(2\pi )=g_{1\!,2;2}/(2\pi )=0.0488$ GHz.

SR also provides us a high-fidelity quantum entangling operation on
charge qubits, assisted by a resonator. This operation gives us a
different way to realize the controlled-phase (c-phase) gate on the
two qubits $q_1$ and $q_2$. Although the amplitude of ROT$_1$ is not
very small, one can see that within a period of ROT$_0$, ROT$_1$
completes two periods accurately, which means that ROT$_1$ does not
change the phase of the state $ \vert 1\rangle _{1}\vert 1\rangle
_{2}\vert 0\rangle _{a}$. In detail, let us assume that $q_{1}$ is
the control qubit and $q_{2}$ is  the target qubit. The initial
state of the system composed of $q_1$, $q_2$, and $R_a$ is prepared
as $\vert \phi\rangle_0=\frac{1}{2}(\vert 0\rangle _{1}\vert
0\rangle _{2}+\vert 0\rangle _{1}\vert 1\rangle _{2}+\vert 1\rangle
_{1}\vert 0\rangle _{2}+\vert 1\rangle _{1}\vert 1\rangle _{2})\vert
0\rangle _{a}$. By  exploiting   the SR on ROT$_0$ and ROT$_1$ and
choosing $g_{0,1;2}t=\pi$, one can get the state of the system
$\vert \phi\rangle_1=\frac{1}{2}(\vert 0\rangle _{1}\vert 0\rangle
_{2}-\vert 0\rangle _{1}\vert 1\rangle _{2}+\vert 1\rangle _{1}\vert
0\rangle _{2}+\vert 1\rangle _{1}\vert 1\rangle _{2})\vert
0\rangle_{a}$. This is just the result of a c-phase gate on $q_1$
and $q_2$. Certainly, by choosing an appropriate transition
frequency of $q_{2}$, one also can get another c-phase gate which
completes the transformation  $\vert \phi\rangle_0 \rightarrow \vert
\phi\rangle_2=\frac{1}{2}(\vert 0\rangle _{1}\vert 0\rangle
_{2}+\vert 0\rangle _{1}\vert 1\rangle _{2}+\vert 1\rangle _{1}\vert
0\rangle _{2}-\vert 1\rangle _{1}\vert 1\rangle _{2})\vert 0\rangle
_{a}$.  From Fig.\ref{fig4}(a), considering the phase error of
indirect interaction between the two qubits and the detune resonance
between $q_2$ and $R_a$, one can get the fidelity of these two
c-phase gates are about $92\%$ within $10.2$ ns.

The quantum entangling operation based on the SR can  also help us
to complete a  single-step controlled-controlled phase (cc-phase)
quantum gate on the three charge qubits $q_1$, $q_2$, and $q_3$ by
using the system shown in Fig.\ref{fig2} except for the resonator
$R_{b}$. Here, $q_1$ and $q_2$ act as  the control qubits, and $q_3$
is the target qubit. The initial state of this system is prepared as
$\vert \Phi\rangle_0=\frac{1}{2\sqrt{2}} (\vert 0\rangle _{1}\vert
0\rangle _{2}\vert 0\rangle _{3}+\vert 0\rangle _{1}\vert 0\rangle
_{2}\vert 1\rangle _{3}+\vert 0\rangle _{1}\vert 1\rangle _{2}\vert
0\rangle _{3}+\vert 0\rangle _{1}\vert 1\rangle _{2}\vert 1\rangle
_{3}+\vert 1\rangle _{1}\vert 0\rangle _{2}\vert 0\rangle _{3}+\vert
1\rangle _{1}\vert 0\rangle _{2}\vert 1\rangle _{3}+\vert 1\rangle
_{1}\vert 1\rangle _{2}\vert 0\rangle _{3}+\vert 1\rangle _{1}\vert
1\rangle _{2}\vert 1\rangle _{3})\vert 0\rangle _{a} $. In this
system, both $q_{1}$ and $q_{2}$ are in the quasi-dispersive regime
with $R_a$, and the transition frequency of $q_{3}$ is adjusted to
be equivalent to that of  $R_a$ when $q_{1}$ and $q_{2}$ are in
their ground states. The QSD transition frequency on $R_a$ becomes
\cite{Alexandre}
\begin{equation}     
\omega'_{a}=\omega
_{a}+\chi_{_1}\,\sigma_{1}^{z}+\chi_{_2}\,\sigma_{2}^{z},
\label{kerrccphase}
\end{equation}
where   $\chi _{_i}=\frac{g_{i}}{\Delta_{i}^{a} }$. By choosing
$g_{0,1;3}t=\pi $, one can realize ROT$_{00}$ (where 00 means the
states of the qubits in the control position are all the ground
states), that is,  $\vert \Phi\rangle_0$ $\rightarrow $ $\vert
\Phi\rangle_1=\frac{1}{2\sqrt{2}}(\vert 0\rangle _{1}\vert 0\rangle
_{2}\vert 0\rangle _{3}-\vert 0\rangle _{1}\vert 0\rangle _{2}\vert
1\rangle _{3}+\vert 0\rangle _{1}\vert 1\rangle _{2}\vert 0\rangle
_{3}+\vert 0\rangle _{1}\vert 1\rangle _{2}\vert 1\rangle _{3}+\vert
1\rangle _{1}\vert 0\rangle _{2}\vert 0\rangle _{3}+\vert 1\rangle
_{1}\vert 0\rangle _{2}\vert 1\rangle _{3}+\vert 1\rangle _{1}\vert
1\rangle _{2}\vert 0\rangle _{3}+\vert 1\rangle _{1}\vert 1\rangle
_{2}\vert 1\rangle _{3})\vert 0\rangle _{a}$. It is just the result
of a cc-phase gate on the three qubits. Fig.\ref{fig4}(b) shows the
probability distributions for the four  quantum Rabi oscillations in
this system. In our simulation, the parameters are chosen as:
$\omega _{r_{a}}/(2\pi )=6.0$ GHz, $\omega _{0\!,1;1}/(2\pi
)=E_{1;1}-E_{0;1}=5.0$ GHz, $\omega _{1\!,2;1}/(2\pi
)=E_{2;1}-E_{1;1}=6.3$ GHz, $\omega _{0\!,1;2}/(2\pi
)=E_{1;2}-E_{0;2}=5.0$ GHz, $\omega _{1\!,2;2}/(2\pi
)=E_{2;2}-E_{1;2}=6.3$ GHz, $\omega _{0\!,1;3}/(2\pi
)=E_{1;3}-E_{0;3}=6.068$ GHz, $\omega _{1\!,2;3}/(2\pi
)=E_{2;3}-E_{1;3}=7.3$ GHz, $g_{0\!,1;1}/(2\pi )=g_{1\!,2;1}/(2\pi
)=0.2$ GHz, $ g_{0\!,1;2}/(2\pi )=g_{1\!,2;2}/(2\pi )=0.2$ GHz, and
$g_{0\!,1;3}/(2\pi )=g_{1\!,2;3}/(2\pi )=0.035$ GHz. The MEAVs of
unwanted quantum Rabi oscillations can be suppressed a lot, and the
fidelity of this cc-phase gate reaches about $86\%$ with $14.8$ ns.

A complex three-qubit gate, such as a Fredkin gate on a three-qubit
system can also be constructed with the quantum entangling operation
based on the SR assisted by the two resonators $R_a$ and $R_b$ in a
simple way, shown in Fig. \ref{fig2}. In this system, $R_{b}$ has a
different transition frequency with $R_a$, and $q_{1}$ couples to
both $R_{a}$ and $R_{b}$ simultaneously in the quasi-dispersive
regime.   Let $q_{2}$ and $q_3$  resonate selectively with $R_a$ and
$R_{b}$ when $q_{1}$ is in the state $\vert1\rangle _{1}$,
respectively, with $g_{0,1;2}^{a}t= g_{0,1;3}^{b}t=1.5\pi$ first
(here $g_{i,j;q}^{k}$ is the coupling strength between the resonator
$k$ and the qubit $q$ in the transition between the energy levels
$\vert i\rangle_q$ and $\vert j\rangle_q$), and then let $q_{2}$ and
$q_3$ be selectively resonant with $R_b$ and $R_{a}$ when $q_{1}$ is
in the state $\vert1\rangle _{1}$, respectively, with
$g_{0,1;3}^{a}t=g_{0,1;2}^{b}t=0.5\pi$, a Fredkin gate can be
realized.

Interestingly,  our gates are significantly faster than previous
proposals in the quasi-dispersive regime in circuit QED
\cite{Jerry,natureletter}. For example, the operation time for a
c-phase gate on two perfect superconducting qubits completed in 2012
is  110 ns with the fidelity 95\% in 2012 \cite{Jerry}. The time for
the cc-phase gate  is 63 ns with the fidelity 85\%
\cite{natureletter}.  By taking the influence from the third levels
of the superconducting qubits into account, the operation time of
our gates is reduced largely.  For example, the operation time of
the c-phase gate on two perfect superconducting qubits is 22 ns with
the fidelity 90\%, about twice of that based on the qubits with the
influence from the third levels.

Our single-step quantum  entangling  gates have some good features.
First, they do not require any kind of drive fields, which
eliminates the limit on the quality of the resonator and does not
increase the temperature of circuit QED, and this factor can protect
the superconducting qubit in circuit QED \cite{Rigetti} and provide
a different way to realize quantum computation.  Second, they are
constructed without using any kind of  the interaction between
qubits (such as Ising-like interaction
 \cite{Nielsen} and Heisenberg-like interaction  \cite{Loss}), similar to the
c-phase gate in Refs. \cite{Jerry}, far different from Refs.
\cite{Michael,AiMinChen}, and our single-step universal quantum
gates can be realized on the non-nearest-neighbor qubits.

Obviously, the  maximal entanglement of superconducting qubit
systems can  be produced effectively with our quantum entangling
operation. On the other hand, the operation requires different
coupling strengths for different qubits with the resonator. This is
not easy to design in a realistic quantum processor.  Luckily, they
can be obtained by using the tunable coupling qubit, which is also
necessary to realize a realistic quantum processor
\cite{RHarries,Srinivasan}. The errors of SR mainly take place with
two points. First, the non-resonance ROT can generate a phase error
with $e^{i\Delta t/2}$. Second, the indirect interaction between
qubits can also generate the unwanted phase error. The suitable
parameters taken to suppress the complex phase error should help us
to get higher fidelity gates. With the stronger coupling of control
qubits, the operation time of our gates can be shortened further.

In conclusion, we have proposed a selective-resonance scheme to
perform a fast quantum entangling operation for quantum logic gates
on superconducting qubits. This approach has many advantages over
previous works.  First,  our quantum entangling  gates are
significantly faster than previous proposals. Second, they do not
require any kind of drive fields. Third, the tunable period relation
between a wanted quantum Rabi oscillation and an unwanted one can
shorten the operation time of the gates, besides the positive
influence from the non-computational third levels of the
superconducting qubits. The principle of our SR   can be generalized
to some other similar cavity QED systems for quantum information
processing.

We would like to thank Prof. Frederick W. Strauch, Prof. Chui-Ping
Yang, and Dr. Qi-Ping Su for helpful discussion. Also, we would like
to thank  Ming-Jie Tao for his help in calculating the probability
distribution. This work is supported by the National Natural Science
Foundation of China under Grant No. 11174039 and NECT-11-0031.

\end{document}